# Alignment of Stakeholder Expectations about User Involvement in Agile Software Development

Jim Buchan[1], Muneera Bano[2], Didar Zowghi[2], Stephen MacDonell[1], Amrita Shinde[1]

[1]*School of Engineering, Computer & Mathematical Sciences*
*Auckland University of Technology, Auckland, New Zealand*
jim.buchan@aut.ac.nz, stephen.macdonell@aut.ac.nz, amritashinde86@gmail.com
[2]*Faculty of Engineering and Information Technology University of Technology Sydney, Australia*
Muneera.Bano@uts.edu.au, Didar.Zowghi@uts.edu.au

**Abstract**

***Context:*** *User involvement is generally considered to contributing to user satisfaction and project success and is central to Agile software development. In theory, the expectations about user involvement, such as the PO's, are quite demanding in this Agile way of working. But what are the expectations seen in practice, and are the expectations of user involvement aligned among the development team and users? Any misalignment could contribute to conflict and miscommunication among stakeholders that may result in ineffective user involvement.* ***Objective:*** *Our aim is to compare and contrast the expectations of two stakeholder groups (software development team, and software users) about user involvement in order to understand the expectations and assess their alignment.* ***Method:*** *We have conducted an exploratory case study of expectations about user involvement in an Agile software development. Qualitative data was collected through interviews to design a novel method for the assessing the alignment of expectations about user involvement by applying Repertory Grids (RG).* ***Results:*** *By aggregating the results from the interviews and RGs, varying degrees of expectation alignments were observed between the development team and user representatives.* ***Conclusion:*** *Alignment of expectations can be assessed in practice using the proposed RG instrument and can reveal misalignment between user roles and activities they participate in Agile software development projects. Although we used RG instrument retrospectively in this study, we posit that it could also be applied from the start of a project, or proactively as a diagnostic tool throughout a project to assess and ensure that expectations are aligned.*

**Keywords:** User involvement; Agile software development; expectations; Repertory Grids; alignment

**General Terms**
Management, Human Factors

## 1. INTRODUCTION

In contemporary software development, frequent user engagement throughout the development process is commonly viewed as good practice, leading to increased development productivity and user satisfaction with the product [1, 2]. Typically, users' involvement relate to their participation in activities related to specifying, elaborating, prioritizing, reviewing and verifying the requirements, as well as testing and verifying developed features [3]. The users involved may include end-users, product owners, project sponsors, subject matter experts, or business analysts [4].

A number of recent studies have provided empirical evidence to support the benefits from such involvement [1, 2, 5-7]. They synthesize the empirical literature and provide empirical evidence that the effects of user participation positively influenced the system development efficiency by providing developers with the domain knowledge they needed. Similarly, a number of studies exemplified by support the positive relationship between user involvement and high client satisfaction.

The effectiveness of UI can vary considerably in different projects and it can be difficult to achieve at times [8]. For example, users may not be fully engaged and provide superficial feedback that results in the need for re-work. This raises concerns about how effective user involvement should be assessed [9]. Some related questions are: 'what activities should the users be involved in to make UI effective?' and 'what users' characteristics may contribute to effective UI?' The perceptions of different stakeholders about the extent and degree of UI can also be misaligned. One of the factors that may influence the effectiveness of UI is the degree of alignment of expectations of different stakeholders regarding UI.

In this paper, we present an empirical study of the stakeholders' expectations about user involvement in Agile software development. We use the term *stakeholder* to refer to everyone involved in the software development project.

In particular, we distinguish between the development team (e.g. business analysts, project manager, testers and developers) and users' representatives (e.g. subject matter expert, product owner).

Our research is motivated by the high-level question: *"How can we assess the alignment of the stakeholders' expectations about user involvement in Agile software development?"*

The major contributions of this research are:

1. An empirical investigation of the alignment between the expectations of Agile software development team and users' representatives about UI.
2. A novel application of the Repertory Grid method for assessing the alignment of expectations of software development team and users.

The rest of this paper is organized as follows: Section 2 provides the background for this research, Section 3 describes the research design, Section 4 presents the results of the case study, Section 5 presents the discussion of the results, Section 6 briefly mentions threats to validity and Section 7, the conclusion and future works.

## 2. BACKGROUND

It is critical for any software product to meet the needs of its users for it to be considered successful. User involvement is a well-known practice in software development and has been acknowledged as an important contributor to project success [10]. This concept can be traced to organizational management research literature, including group problem solving, interpersonal communication and individual motivation [4]. For four decades researchers have investigated this phenomenon and among many results obtained, they reported that effective user involvement can lead to system success [1, 11, 12].

Research on Agile software development has increased rapidly in the last decade [13]. Active and continuous participation of users throughout the project is at the core of Agile software development approach [14, 15]. Agile approaches such as Scrum and XP, emphasize daily interactions between users' representative (the PO) and the development team [16]. The frequent interaction between users' representatives and the development team is aimed at strengthening a collaborative partnership. This practice ensures that users have a meaningful influence on decisions about system features and the work priorities. The benefits of UI have long been recognized in research literature [12]. These benefits include: user satisfaction [6, 17], improved communication [18], improved software quality [19], improved quality of design decisions [20], and facilitating change [21] to name but a few.

Frequent user involvement, does not imply effective involvement. Achieving effective user involvement is complex and multifaceted because many factors have been recognized to play important roles in achieving the desired benefits and goals of UI. For example, identification of the right representative of users is essential. There are different stakeholders from the user group whose influence on the quality of user involvement, directly or indirectly, is crucial for effectiveness. Each user type has a different level of influence on the project depending on their level of authority in their organization, their knowledge, expertise and experience. Based on this variation, a user can be classified as: end user, user representative, Subject Matter Expert (SME), Product Owner (PO), or proxy to client [22].

User involvement in one activity of software development is said to influence the level of involvement in the subsequent activities [23]. Kujala [24] stressed that involving users in early stages is more beneficial. Others claimed that after effective involvement of users during requirements determination further involvement may not be necessary at later stages [25]. However, unlike previous claims made in plan based software development, in an Agile approach, UI is conceived as a continuum. Hence it is essential to differentiate the extent and the degree of UI in different activities of the Agile software development process, in order to assess the effectiveness of UI.

There are two distinct views of the extent of effective UI in various activities of software development: 1) the perception of the users' representatives about their involvement, and 2) the perception of the development team about the involvement of the users' representatives. If the expectations of the development team about user involvement differ considerably from that of the users' representatives, then this misalignment can have a significant impact over the effectiveness of UI and may impede productivity. For example, if a Product Owner or Subject Matter Expert has much lower expectations about their time commitment or required technical knowledge needed for effective involvement in a project, compared to the development team's expectations (e.g. Developer, Tester, Business Analyst), then this may lead to conflict, confusion and higher risk of project failure.

As evidenced in a recent Systematic Review [12], the alignment of stakeholder expectations about user involvement in Agile software development has not received much attention in the research literature. This paucity of knowledge about alignment of expectation has motivated us to investigate this important and complex phenomenon in practice. Our interest in the research reported in this paper was also partially motivated from extensive engagement with many Agile teams in practice, in particular, informal conversations with several PO's, where the POs expressed their surprise at the high level of involvement that the development team expected of them. This was complemented by informal observation of the progress of some development teams being held up by lack of availability of user expertise at the time it was needed, and slow or poor quality feedback.

## 3. RESEARCH DESIGN

The aim of this research is to empirically investigate the expectations of various stakeholders in the context of user involvement in agile software development. We conducted a case study in order to gain in-depth knowledge of the complexity of user involvement in practice by focusing on its various inter-connected constituents such as user involvement in various SDLC activities, characteristics of user roles, expectations of users as well as the development teams about user involvement. The data collection strategies included conducting a literature review, interview study and repertory grids.

### 3.1 Case study

The case study was exploratory and interpretive in nature to compare and contrast the expectations of various stakeholders about user involvement. Case studies are widely acknowledged for their applicability to real life contexts or industrial settings, and their flexibility, in which the researcher employs multiple methods for data collections [26]. An interpretive qualitative research inquiry is well suited to our research as it is based on the assumption that the reality can be understood through the social construction of language, consciousness and shared meanings [26].

The selected organization (ABC)[1] is a medium-sized company in the Insurance/finance sector and is situated in Auckland, New Zealand. They use Agile software development methodology with active user involvement. ABC has been using an Agile approach to software development for over 5 years and is relatively mature in its use of Agile methods The IT department is responsible for maintaining several internal systems as well as outward facing systems that support their main financial/insurance services. The participants reported in this paper were all part of two development teams, one team developing a system for internal end-users, and the other team developing a system for external end-users. From the first team, 2 DEVs, 1 BA, 1 PM, 1 T were interviewed, and from the second team, 2 DEVs, 1 BA, 1 PM, 1 PO and 1 SME were interviewed. All participants had between 5 and 12 years' experience in their roles, and had been with the organization between 1 and 5 years. The PO also had a role as a manager with the organization. The SME was a former end-user external to the company.

#### 3.1.1 Unit of analysis

As this study investigates the alignment of expectations based on the perceptions of the people whose functions fall under roles of either the user or development teams, the unit of analysis considered was the 'role' of the interviewee. This was later used in the repertory grids to analyze the inter-role alignment of expectations.

#### 3.1.2 Data collection

Our data collection strategy involved multiple sources to strengthen the validity and reliability.

**Literature review:** The first stage of the research began with a thorough literature review of the concepts and theories surrounding user involvement and agile methodology of software development. User involvement in software development is an area of research that has been explored extensively since 1970s [1, 4, 11, 12]. The findings from literature review were used to formulate and design an instrument for semi-structured interviews. Later the findings from the literature review combined with the results of the interviews were used to define the elements and constructs of repertory grids (see Figure 1).

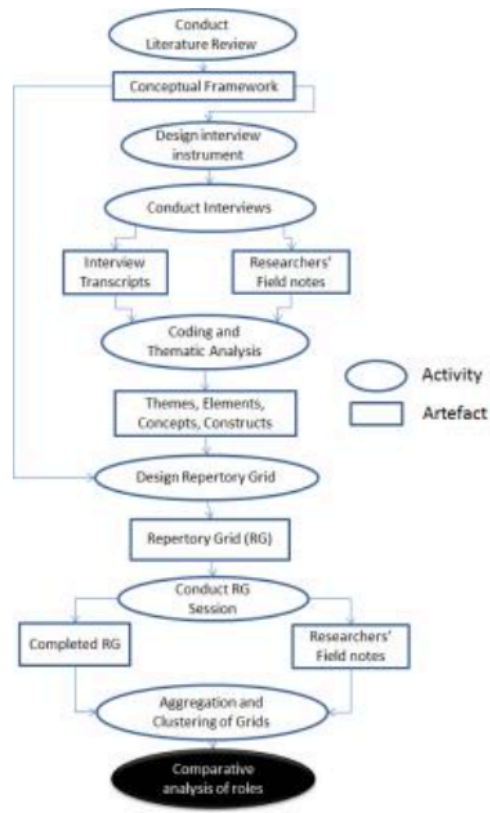

Figure 1. Research Design

**Semi-Structured Interviews:** As this research aims to understand the practitioners' expectations of user involvement of various roles in context of agile software development, we collected data about their experience and expectation from both users and developers.

We conducted 11 face-to-face semi-structured interviews by asking open-ended questions (see Appendix A) each lasting one hour on average. The participants included 4 Developers (DEV), 2 Project Managers (PM), 2 Business Analysts (BA), 1 Tester (T) and 2 members of users group, a Product Owner (PO) and Subject Matter Expert (SME). The interviewees were asked to respond and explain while retrospectively reflecting on their experiences by describing the situations they have encountered in context of user involvement in their currently on-going projects or recently finished projects. All the interviews were completed over the period of one week.

The interviews were audio-recorded and partially transcribed by the researchers to extract only the relevant information from audio recordings. Two of the researchers were present during all interviews and took field notes as necessary. The partial transcripts and field notes were later coded and thematically analyzed.

**Interview data analysis:** Thematic analysis in the form of template analysis was applied on the data collected from the interviews. Template analysis is well suited for comparing perceptions of different groups within specific context [27]. Template analysis enables the development of conceptual themes that fall into bordering groups eventually enabling the identification of master themes and their subsidiary constituent themes.

---

[1] The name of the organization and the practitioners will not be disclosed as per Ethical Research agreement from Auckland University of Technology.

**Repertory grids:** The secondary element to the data collection method employed was the use of the Repertory Grid technique based on the Personal Construct Psychology [28]. RG is a form of cognitive modeling and we utilized it in our research to represent shared understanding among members of groups of subjects, in our case users and development team. The basic components of RG are elements (columns) and constructs (rows) which form the grid. The elements are the objects that are the focus of the study and the constructs are the ideas that the participants hold about elements [29]. From analysis of literature and interviews, we defined the elements and constructs for our RG that represented various roles and activities of user involvement in Agile software development.

We applied the RG for two contexts: expectations of user involvement in various SDLC activities, and expectations of user characteristics for effective involvement. The user roles PO and SME were defined as the *elements* for RG and were fixed for both grids. The RGs were conducted with these user roles, as well as the BA, PM, DEV, and T from the development team. The different activities were investigated (from literature and interview data analysis) to focus on development teams' preferred level of user involvement, for each activity for each role. These activities were defined as *constructs* for the *nature of involvement in* RG to enable comparison between the perspectives of different roles. Similarly, the expected characteristics of the user roles for effective UI were defined as the *constructs* for the *user characteristics* RG to compare expectations of the roles.

The difference in the level of expectations was recorded on a 7-point Likert scale where 1 represents *very low involvement* and 7 represents *very high involvement.* Scale of 7 was selected to provide the participants with more scope for expressiveness. The participants were asked to evaluate the construct for preferred level of involvement and level of importance against this scale.

**RG data analysis:** The grids completed by participants were aggregated based on their project roles using Frequency Distribution analysis. Treating the Likert scale as an interval scale, the aggregated frequency values for each cell were calculated from the weighted average of scores for each role, converting the 7-point Likert scale to a 7–point scale from -1 to 1, with zero as the neutral point. These were then classified into 5 categories based on one standard deviation (0.7) around the neutral point, labelled Very Low (VL) to Very High (VH) as shown in Table 1.

The aggregated grids for each role (PM, T, DEV, BA) were cross compared with the user roles (PO, SME) to find patterns of similarities and differences of expectations to identify significant alignment or misalignment of expectations among these roles.

***Protocol for Conducting the Repertory Grid:*** To ensure participants interpreted the high-level activities in the RG

Table 1. RG data analysis

| Aggregated Frequency (F) | Category |
|---|---|
| F > 0.7 | Very High (VH) |
| 0 < F < 0.7 | High (H) |
| 0 | Neutral |
| - 0.7 < F < 0 | Low (L) |
| F < 0.7 | Very Low (VL) |

consistently the grid elements and constructs were explained to each participant and any questions they had were answered. To further reduce this potential threat to validity, participants were asked to "think out loud" as they completed the grid, providing further opportunity to address any misunderstandings. The participant was then asked to answer the question:

*For each of the different client roles what would you expect the involvement of each of them to be for the different activities, over the whole project?*

The 7-point Likert scale for representing their perception of the level of involvement on the grid was then explained and they were asked to vocalize their reasoning as they filled in each grid cell. A similar protocol was followed for the second grid, with the question:

*For each of the different roles how would you describe the importance of each of the following user characteristics to their high-quality involvement?*

## 4. RESULTS

### 4.1 Exploratory Interviews

The overall purpose of the interviews was to explore different facets of users' involvement in Agile software development. The scope of this paper is limited to the analysis of the interview questions that are directly related to the design of RG.

#### 4.1.1 User Roles
To investigate the expectations of user involvement, the question of which user roles to focus on was firstly investigated. One of the aims of the interviews was to validate the user roles identified in the literature review. Participants in the interviews were asked to identify who they consider to be the main users' representative that should be involved in the project. The roles mentioned by the participants were mapped against those reported in the literature and thus classified into five main user roles (End Users, Product Owners, Subject Matter Experts (SMEs), Client Managers, Client Proxies).

The interviews confirmed the central importance of the involvement of the PO as a user role in the case organization. It was also identified that the SME had high levels of involvement in the projects. Although our interview results included all the roles mentioned above and subsequently we used all roles in RG design, in this paper, we focus only on PO and SME. The reason being that only PO and SME were available to participate in RG session within the time frame of the data collection.

An SME is generally described as someone with specialized domain expertise. The SME participating in this study was someone who was previously an experienced user of the product and had excellent product and process knowledge as well as insight into end users' challenges. The PO was identified as an important user role by every interviewee and was described as someone who consults widely with other users' representative and has the authority to make decisions about feature priority. Interviewees noted that it was expected that the PO is highly available throughout the project for explanations, verification and decisions.

### 4.1.2 User Activities
The main categories of user activities identified by thematic analysis of the interviews are presented and defined briefly in Table 2. The constructs (rows) of the first RG comprised these user activities.

### 4.1.3 Desired User Characteristics
The main categories of user characteristics identified by thematic analysis of the interviews are presented and explained briefly in Table 3. The constructs (rows) of the second Repertory Grid comprised these user characteristics.

### 4.2 Repertory Grid Analysis
This section presents the results of aggregating expectations by role and coding these from Very Low (VL), Low (L), Neutral (N), High (H), and Very High (VH).

Firstly, the expectations of the PO's and SME's levels of involvement in user activities are presented in Table 4 and Table 5 respectively, followed by the expectations of the characteristics of the PO and SME in Table 6 and Table 7. In Table 4 the PO column presents the PO's expectations of his involvement. The DEV, BA, PM, and T columns represent their expectations of the PO's involvement, respectively. In Table 5 the SME column presents the SME's expectations of his involvement in the user activities, and the next four columns present the expectations of the corresponding development team roles regarding the SME's involvement in these user activities.

The PO's and SME's expectations about their own characteristics, if they are to contribute effectively, are summarized in the first column of Table 6 and 7, respectively. The DEV, BA, PM and T columns in Table 6 represent the expectations of these roles regarding the PO's characteristics, and the corresponding columns in Table 7 represent their expectations of the SME's characteristics.

To facilitate easy comparison of expectations on a role-by-role basis, the expectations are presented as heat maps with the shading of each cell depicting a level of misalignment of each development team role's expectations with the user role's expectations. Darker grey indicates higher levels of misalignment with the user role's expectations. Patterns of alignment and misalignment are then identified and discussed for each table and overall.

### 4.2.1 Expectations of PO's involvement in User Activities
Table 4 suggests that the PO expected to have a high involvement in activities related to decisions about the product features including writing, eliciting, prioritizing and validating requirements (A1-A4). These expectations align well with the team roles' expectations generally. The exception is the misalignment of the BA's expectations that the PO would have almost no involvement in user story writing (A3) and low involvement in requirements clarification (A4). Based on what the BAs said as they were completing this grid cell, this misalignment seems to have been based on the BAs' broader view of their own responsibilities in these activities. This included writing user stories based on a variety of requirements sources, and gathering information to clarify requirements from a variety of information resources as well as communicating these to the team.

Table 2. Activities User may be Involved in

| | |
|---|---|
| A1. Decision Making about product and/or process | Decisions about which features to include, navigation, look and feel of the product and user interface, and the overall workflow of the product. |
| A2. Requirements Elicitation | Communicating with relevant stakeholders (e.g. end-users) to elicit and negotiate their needs and requirements and present these to the development team. |
| A3. User stories | Writing user stories including story creation and editing. |
| A4. Clarification of Requirements | Explaining context and details and answering questions about specific user stories. |
| A5. Verification of Functionality (e.g. UAT) | Evaluating and reviewing completed functionality and features. |
| A6. Co-development | Creating functionality by coding at some level. |
| A7. Cost Negotiation | Negotiating budget and cost being involved in project cost estimation. |
| A8. Time Frame/Schedule of Project | Prioritising features and work and updating release plans. |
| A9. Choice of Technology | Being involved in selecting the technology related to building or delivering the product |

Table 3. User Characteristics for Effective Involvement

| | |
|---|---|
| C1. Time Investment per week | The number of hours the user expected to commit to the project per weeks |
| C2. Ability to Articulate | The user's ability to communicate clearly their knowledge and understanding, and uncover assumptions. |
| C3. Desire to Participate | The user's motivation and willingness to participate in all project related activities they are needed in. |
| C4. Availability | Available for ad hoc and planned activities |
| C5. Authority and level of Influence | The user has authority to make appropriate decisions and represent others. |
| C6. Experience with project process | Is experienced with Agile approach of project management |
| C7. Technical knowledge | Has knowledge of the technology and tools used for the product. |
| C8. Business process and product knowledge | Has knowledge of the business process context and of the features of the existing product. |

There is a high level of misalignment of expectation when it comes to the PO's involvement in verification of feature functionality (A5). The development team roles, apart from the BA, are expecting the PO to have very high involvement in this activity. In contrast, both the PO and BA roles are expecting the PO to have very little contribution to this activity (A5). While completing this part of the Grid, the PO explained that the SME, as a former power user of the product, would be better suited to this and hence the PO needed little involvement.

The first column of Table 4 shows that the PO did not expect to be involved in co-developing the code and choice of technology, and this aligns with the expectations of the development team roles. While filling in the Grid, the PM explained that co-development could include involvement in less technical activities with developers such as planning poker, prioritization meetings and acceptance test creation, hence the higher expectation of the PM.

Table 4 shows that the PO expected to have some involvement in project cost negotiation (A7), but not high involvement. This could be a consequence of the other role of the PO as a manager, which he described as involving setting and justifying project budget variations. The main misalignment of expectations about the PO's involvement in A7, is seen in the T column of Table 4 where the Tester expected the PO to have a very high involvement. The Tester was new to the team and the Agile approach and she emphasized that the PO is more of a manager, which could explain this high expectation.

Table 4 shows that the PO expected to have some involvement in project cost negotiation (A7), but not high involvement. This could be a consequence of the other role of the PO as a manager, which he described as involving setting and justifying project budget variations. The main misalignment of expectations about the PO's involvement in A7, is seen in the T column of Table 4 where the Tester expected the PO to have a very high involvement. The Tester was new to the team and the Agile approach and she emphasized that the PO is more of a manager, which could explain this high expectation.

Table 4. PO Activities: Team's and PO's expectations of the PO's involvement

|  | PO | DEV | BA | PM | T |
|---|---|---|---|---|---|
| A1. Decision Making about product and/or | VH | H | N | VH | VH |
| A2. Requirements Elicitation | H | H | H | VH | VH |
| A3. User stories | H | H | VL | VH | N |
| A4. Clarification of | H | VH | L | VH | H |
| A5. Verification of Functionality (e.g. UAT) | VL | VH | L | VH | VH |
| A6. Co-development | VL | VL | VL | N | VL |
| A7. Cost Negotiation | N | L | N | N | VH |
| A8. Time Frame/Schedule of Project | H | H | L | H | H |
| A9. Choice of Technology | L | L | VL | VL | L |

From the first column of Table 4, the PO expected to have high involvement in work scheduling (A8). From what was said during the Grid completion, this related to the PO's involvement with user story prioritization. The Table shows that this same expectation was shared by the Development Team roles, apart from the BA who said he expected the PM to have scheduling responsibility.

The overall pattern seen in Table 4 is one of general alignment of expectations between the PO and the Development Team, as well as between roles within the Development Team. The most significant misalignment of expectation relates to the PO's involvement in functional verification (A5).

### 4.2.2 Expectations of SME involvement in User Activities

The SME's expectations of involvement in different user activities are generally low to neutral, as seen in the first column of Table 5. During the Grid completion, the SME explained this by describing his uncertainty about the role he would play in this project presumably because of his lack of experience with Agile and this project. The development team members were unaware of the SME's uncertainty and the subsequent columns show they generally have an expectation of much higher involvement of the SME in the user activities.

Table 5 shows that the DEV, and T expected the SME to have high involvement with activities A1 to A5, in contrast to the SME's own expectation. Table 4 also shows that the BA's expectations of the SME's involvement in A1 to A3 are misaligned with the high expectations of the DEV and T, and more aligned with the SME's expectations of low involvement in these activities.

Table 5. SME Activities: Team's and SME's expectations of the SME's involvement

|  | SME | DEV | BA | PM | T |
|---|---|---|---|---|---|
| A1. Decision Making about product and/or process | N | VH | L | VH | H |
| A2. Requirements Elicitation | N | VH | L | VH | H |
| A3. User stories | L | VH | L | N | H |
| A4. Clarification of Requirements | N | VH | H | VH | VH |
| A5. Verification of Functionality (e.g. UAT) | N | VH | H | L | VH |
| A6. Co-development | VL | L | VL | VL | VL |
| A7. Cost Negotiation | VL | L | VL | VL | N |
| A8. Time Frame/Schedule of Project | N | N | L | VL | N |
| A9. Choice of Technology | L | L | L | VL | L |

A similar pattern is seen with A5 in Table 5, where the DEV, BA and T have high expectations of the SME's involvement in this activity, misaligned with the SME's and PM's low expectations. Rows A6 to A9 depict a general alignment of expectation that the SME will not have high involvement in these activities.

Table 5 shows an overall the pattern of misalignment of the SME's neutral expectations and the Development Teams higher expectations in terms of the SME's involvement in requirements- and product-related activities A1 to A5. In contrast, both the development team and the SME share a low expectation of involvement in the technical and project-related activities A6-A9.

### 4.2.3 Comparison of PO and SME User Activities

In Table 4 the BA and PO had a low expectation of the PO's involvement in A5 and Table 5 shows their expectation of the SME involvement in A5 is high, suggesting BA and PO were expecting A5 mainly to be the responsibility of the SME. From Table 5 it is also apparent that the rest of the development team roles, apart from the PM, expected both the PO and SME to have high involvement in A5. Looking at the PM's expectation in Table 3 it seems that the PM expected most of the functional verification to be done by the PO.

**4.2.4 Expectations of PO Characteristics**
From the PO column of Table 6 it is apparent that the PO expected to commit significant time to this project (C1) and to be highly available to the team (C4). These expectations align with the expectations of the DEV, PM and T for C1 and C4. The BA column, suggests that the BA expected less of a time commitment (C1) and lower availability (C4) of the PO.

The C2 row of Table 6 shows that the PO and Development Team were all in agreement that the ability to articulate ideas well is a highly desirable characteristic for a PO to be effective. Similarly, rows C3, C5 and C8 show a strong alignment of expectations related to the importance of these PO characteristics. There is also general agreement on the expectation that technical knowledge (C3 row) has low importance for effective involvement of the PO.

Analysis of the C6 column in Table 6 shows some misalignment in expectations of the PO's project process experience for effective PO involvement. The PO and PM both consider this of low importance compared to the high importance placed on C6 by the DEV, BA and T.

Table 6. PO Characteristics: Team's and PO's expectations of PO's characteristics

|  | PO | DEV | BA | PM | T |
|---|---|---|---|---|---|
| C1. Time Investment per week | VH | H | N | VH | VH |
| C2. Ability to Articulate | VH | VH | H | VH | VH |
| C3. Desire to Participate | VH | VH | H | VH | VH |
| C4. Availability | VH | H | N | VH | H |
| C5. Authority and level of Influence | VH | VH | VH | VH | VH |
| C6. Experience with project process | L | H | H | L | H |
| C7. Technical knowledge | L | L | L | L | L |
| C8. Business process and product knowledge | VH | H | H | VH | VH |

**4.2.5 Expectations of SME Characteristics**
The SME column of Table 7 shows that the SME expects all characteristics to be high or very high for effective SME involvement, except project process experience (C6), which has a lower (neutral) importance. Rows C1 to C8 in Table 7 show that the development team's expectations are generally in very strong alignment with the SME's expectations, apart from C6 and C7.

Considering the C6 row in Table 7, the SME and BA expect that the SME should have a high level of technical knowledge for effective SME involvement, whereas the DEV, BA and T expect the SME's technical expertise to be of far less importance for effective SME involvement.

An analysis of the C6 row in Table 7 shows that the DEV, BA and T put high importance on the project experience of the SME, whereas the SME considers it less important and the PM considers it of even lower importance. The SME considered it more important for themselves to have project process experience, and even more so technical knowledge, compared to the importance placed on these by the PO.

Table 7. SME Characteristics: Team's and SME's expectations of SME's characteristics

|  | SME | DEV | BA | PM | T |
|---|---|---|---|---|---|
| C1. Time Investment per | VH | VH | VH | VH | VH |
| C2. Ability to Articulate | VH | VH | VH | VH | VH |
| C3. Desire to Participate | H | VH | VH | VH | VH |
| C4. Availability | H | VH | VH | VH | H |
| C5. Authority and level of Influence | H | H | H | VH | H |
| C6. Experience with project process | N | H | H | L | H |
| C7. Technical knowledge | H | L | H | L | L |
| C8. Business process and product knowledge | VH | VH | VH | VH | VH |

**4.2.6 Comparison of PO and SME Characteristics**

Comparing the first columns of Tables 6 and 7 suggests that both SME and PO considered it very important to commit a high level of time per week and to be articulate (C1 and C2). Generally, the SME thought the characteristics C3-C5 were slightly less important for an SME than the PO thought they were for a PO to be involved effectively.

Further comparison of Tables 6 and 7 shows that the development team's expectations of both the SME's and PO's characteristics are very similar overall. One exception to this pattern is that the PM's considers that project process experience (C6) has low importance for both the PO and SME, whereas the DEV, BA and T consider this characteristic to have high importance for both the PO and SME.

**4.2.7 Comparison of User Activities with User Characteristics for the PO and SME**
Comparison of the expected User Activities results (Tables 4 and 5) with the results of expected User Characteristics (Tables 6 and 7), provide some insights into what user knowledge and commitment the development team expects for the activities they expect the SME and PO to be involved in.

From Table 4, the development team generally expects the PO to have high involvement in activities related to the product look and feel (A1), and requirements-related activities (A2-A4) as well as verification of functionality (A5) and influencing work scheduling (A8). Comparison with Table 5 reveals that to be involved in this way the development team expect the PO to have little technical knowledge (C7) but be very knowledgeable about the product and business process (C8) as well as the project management process (C6). They also expect that involvement in these activities will require a high time commitment and availability (C1 and C3), high motivation to participate (C2) and will need appropriate authority (C5).

For the SME, Table 5 shows that the development team has an overall expectation that the SME will mainly be involved in requirements-related activities A2 to A4, decisions about product look and feel (A1) and functional verification (A5). To be involved in this way the development team expects the SME to have very similar characteristics to the PO (Table 7).

# 5. DISCUSSION

On the surface there seemed to be a misalignment of expectations between some user roles and the development team regarding user involvement.

Our results in this case study revealed that the development team's expectations of both the PO's and SME's levels of involvement in user activities as well as user characteristics is mostly aligned with the PO's and SME's own expectations, with just a few exceptions. Therefore, the reasonable level of alignment of expectations observed was somewhat unexpected. The reasons for this may lie in the context of the teams and the projects investigated and the following are hypothesized as possible explanations:

1. *A mature Agile team.* The team members, apart from the SME, were very familiar with the reasonably stable Agile development practices used in the organization and had previously worked together, with shared experiences to draw on. They had ample opportunity to share understanding of expectations.

2. *The teams were co-located and the projects were developed with in-house resources.* It was relatively easy for all team members to meet and share expectations with each other and the user stakeholders. This increased the opportunity for (re)alignment of expectations.

3. *Experienced PO.* The PO had been in the role within the organization several times and had experienced the high involvement needed of the PO throughout the software development process, and was therefore expecting it.

Within this general pattern of alignment, there are some areas of misalignment, which are now discussed.

## 5.1 Misalignments of Expectations of User Involvement

There is a strong misalignment of expectations regarding the PO's and SME's involvement in activities related to verifying new functionality (A5 in Tables 4 and 5), where both the PO and SME have low expectations of their own involvement, in contrast to the development team's high expectation of the SME's and PO's involvement in A5. This misalignment increases the risk of a hold up when it comes to functional verification and User Acceptance Testing, with PO and SME not expecting to be involved and the team assuming it is their responsibility.

As seen Table 5, (rows A1 to A5), the SME's low expectation and misalignment with the high expectation of development team extends to all the requirements-related activities (A1-A4). The SME's low expectation of involvement in A1-A5 is likely to be a by-product of his general lack of clarity in his responsibilities in this project, and fits his overall expectation of having no "high" involvements in any activity, as seen in Table 5. This SME's uncertainty in his role seems to have arisen from a number of circumstantial factors: no previous experience in this role; the SME's lack of experience with how the team worked including the Agile process used by the team; limited previous shared experience working with team members, and even then, his role was an end-user; the team's tacit assumption that he was briefed on his role before being seconded full-time to the team.

The team's high expectations of the SME's involvement are reasonable since the SME was known to some of the team members as a "power user" and is likely to be perceived as representing "end-users". Some planned mechanism to on-board new team members, or having a planned stakeholder meeting about roles and responsibilities prior to starting development, would have surfaced the misaligned expectations and obviated this situation. The PO's expectation of low involvement in A5 is likely to be mainly influenced by the fact that someone else in the team, the SME, was well-suited to tasks in A5, and the PO was already time-challenged with his primary role in the organization as a manager. The team's high expectations of the PO's involvement in A5 may have come from previous experiences in other projects, where a PO was highly involved in functional verification activities. Again, a short, planned meeting about user stakeholder responsibilities to make these assumptions explicit, incorporated into the pre-sprint process, may have addressed this.

Another misalignment of expectations is the BA's much lower expectation than the other team members of the PO's involvement in user activities A1, A3, A4, A5 and A8 (Table 4). This misalignment between the BA and the rest of the team could result in the team over-relying on the BA as a proxy PO, rather than involving the PO in all matters. This could increase the risk of missing or misunderstood requirements, as well as potentially reduce the engagement of the PO if he perceives the BA as fulfilling his responsibilities.

The BA's expectations of the SME's involvement in activities A1 to A3 (Table 5) were also misaligned with the expectations other development team members. The BA seems to have the view that the main involvement of the SME will be clarifying requirements (A4) and verifying functionality (A5). The BA clearly expected mainly the PO to elicit requirements (A2 in Table 4). In the interview the BAs gave the impression that story writing (A2) and managing story cards were the responsibility of the BA. This misalignment may have limited consequences for the project outcome since the requirements activities were collaborative throughout the project, with numerous opportunities for the SME and PO to influence requirements in activities A1 to A4.

## 5.2 Misalignments of Expectations of User Characteristics

The expectations of both the SME and PO were that experience with the project process was not very important for them to have effective involvement (C6 in Tables 6 and 7). This contrasted with the development team who generally thought that this was an important user characteristic for both user roles. For the PO, the low importance of this characteristic probably comes his familiarity with the process from previous experience and should have little consequence. For the SME, this low expectation of knowledge of the project process may be more problematic. They may be underestimating the learning curve needed to adapt to the Agile way of working used throughout the project, and could be a source of stress and confusion for the SME. If the SME's on-boarding program had included some training in the Agile process used, this could have alleviated this risk.

Another misalignment between the SME and development team relates to technical knowledge (C7 in Table 7). This seems to have arisen from different interpretations of the C7 construct, as identified from the participants' thinking out loud during Grid completion. The SME (and BA) interpreted C7 as detailed knowledge about the product technicalities and its business context, whereas the rest of the development team viewed it as knowledge about the technologies used to deliver the product.

This research suggests some possible actions that promote the alignment of expectations of users' involvement. This includes, a transparent process for selection of user representatives which includes input from the development team; a brief team meeting during team formation to on-board users and clarify and align expectations of involvement and why. A simplified form of the RG instrument used in this research could support this early alignment, as well as being useful as a diagnostic instrument during software development, if misalignment of expectations is suspected.

## 6. THREAT TO VALIDITY

The main threat to the validity of the results of a case study are considered to be the context-specific variables of the projects (including organization size and type, project complexity, software development methodology, number and diversity of interview subjects, and size of data set), which makes it difficult to generalize the findings that emerged from the data analysis [30]. With regards to the qualitative analysis in a constructivist paradigm of inquiry, it is impossible to claim absolute exactness of the results free from researchers' biases.

## 7. CONCLUSION AND FUTURE WORK

In this paper, we have presented the result of an exploratory case study to investigate the alignment between the expectations of Agile software development team and users about UI. Qualitative data was collected through interviews to design a novel method of applying Repertory Grids for the assessment of the alignment of expectations about UI. Our analysis from aggregating the results from the interviews and RG revealed varying degrees of expectation alignments between the development team and users' representatives. Application of the RG technique to compare expectations about user involvement was engaging for the participants and relatively straight forward for the researchers. The initial interview and literature review generated a large number of constructs about UI that proved useful in comparing expectations. The process of conducting RG provided an intuitive structure to elicit participants' perceptions that was easy to understand and follow. The think-out-loud aspect of the RG protocol we introduced provided extra insights into the reasoning of the participants and their interpretation of the constructs.

This method can reveal mismatches between user roles and activities they participate in for Agile software development projects. Although we used RG instrument retrospectively in this study, we posit that it could also be applied from the start of a project, or proactively as a diagnostic tool throughout a project to examine and ensure expectations are aligned. The use of RG at the start of a project shows good potential as the trigger to have conversations about the expectations of roles and involvement and surface misalignments, before they become critical. In addition, the RG instrument could easily be used as part of a team diagnostic toolbox to diagnose misalignments of expectations as the root cause of team problems.

The findings from this paper contribute to the body of empirically-based knowledge related to Agile software development practices. This paper deepens the understanding of factors related to alignment of expectation of user involvement, provides empirical evidence for the strength of the alignment of those expectations in practice across team roles. In addition the study contributes to a clear and consistent conceptualisation of the meaning of "effective user involvement". Practitioners will benefit from a deeper understanding and awareness of the differences and similarities of expectations of various development roles and user roles with respect to user involvement. This knowledge should encourage sensitivity to differences and promote discussion and effort to align expectations. A clear and detailed conceptualisation of the meaning of "effective" user involvement provides consistent terminology to discuss the issues related to high quality involvement and what can be expected in what areas. This is useful to both practitioners and researchers interested in exploring this area. It clarifies the meaning of the goal of high quality user involvement and could provide the basis for a future quality metric.

The deeper understanding of factors related to alignment and misalignment of expectations and their influence on the quality of user involvement, as well as clarity in the goal and barriers to achieving that goal, can also provide guidance in the design of future techniques and tools to support the alignment of expectations high quality user involvement.

## REFERENCES

1. He, J. and W.R. King, The role of user participation in information systems development: implications from a meta- analysis. Journal of Management Information Systems, 2008. 25(1): p. 301-331.
2. Hsu, J.S.-C., et al., Users as knowledge co-producers in the information system development project. International Journal of Project Management, 2012. 30(1): p. 27-36.
3. Iivari, N., "Constructing the users" in open source software development: An interpretive case study of user participation. Information Technology & People, 2009. 22(2): p. 132-156.
4. Ives, B. and M.H. Olson, User involvement and MIS success: A review of research. Management science, 1984. 30(5): p. 586-603.
5. Procaccino, J.D. and J.M. Verner, Software developers' views of end-users and project success. Communications of the ACM, 2009. 52(5): p. 113-116.
6. McKeen, J.D., T. Guimaraes, and J.C. Wetherbe, The relationship between user participation and user satisfaction: an investigation of four contingency factors. MIS quarterly, 1994: p. 427-451.
7. Thalen, J. and M. van der Voort, Facilitating user involvement in product design through virtual reality. 2012: INTECH Open Access Publisher.
8. Zowghi, D., F. da Rimini, and M. Bano. Problems and challenges of user involvement in software development: an empirical study. in Proceedings of the 19th EASE. 2015.

9. Kujala, S., Effective user involvement in product development by improving the analysis of user needs. Behaviour & Information Technology, 2008. 27(6): p. 457-473.
10. Bano, M. and D. Zowghi. User involvement in software development and system success: a systematic literature review. in Proceedings of the 17th EASE 2013.
11. Abelein, U. and B. Paech, Understanding the influence of user participation and involvement on system success–A systematic mapping study. Empirical Software Engineering, 2015. 20(1): p. 28-81.
12. Bano, M. and D. Zowghi, A systematic review on the relationship between user involvement and system success. Information and Software Technology, 2015. 58: p. 148-169.
13. Hoda, R., et al., Systematic literature reviews in agile software development: A tertiary study. Information and Software Technology, 2017.
14. Brhel, M., et al., Exploring principles of user-centered agile software development: A literature review. Information and Software Technology, 2015. 61: p. 163-181.
15. Williams, L. and A. Cockburn, Guest Editors' Introduction: Agile Software Development: It's about Feedback and Change. Computer, 2003. 36(6): p. 39-43.
16. Highsmith, J.A., Agile software development ecosystems. Vol. 13. 2002: Addison-Wesley Professional.
17. Bano, M., D. Zowghi, and F. da Rimini, User Satisfaction and System Success: An Empirical Exploration of User Involvement in Software Development. Empirical Software Engineering, 2016.
18. Korkala, M., M. Pikkarainen, and K. Conboy. Distributed agile development: A case study of customer communication challenges. in International Conference on Agile Processes and Extreme Programming in Software Engineering. 2009. Springer.
19. Hunton, J.E. and J.D. Beeler, Effects of user participation in systems development: a longitudinal field experiment. Mis Quarterly, 1997: p. 359-388.
20. Pekkola, S., N. Kaarilahti, and P. Pohjola. Towards formalised end-user participation in information systems development process: bridging the gap between participatory design and ISD methodologies. in Proceedings of the ninth conference on Participatory design: Expanding boundaries in design-Volume 1. 2006. ACM.
21. Hunton, J.E., Involving information system users in defining system requirements: The influence of procedural justice perceptions on user attitudes and performance. Decision Sciences, 1996. 27(4): p. 647-671.
22. Terry, J. and C. Standing, The value of user participation in e-commerce systems development. Informing Science: International Journal of an Emerging Transdiscipline, 2004. 7: p. 31-45.
23. Palanisamy, R., User involvement in information systems planning leads to strategic success: an empirical study. Journal of Services Research, 2001. 1(2): p. 125.
24. Kujala, S., User involvement: a review of the benefits and challenges. Behaviour & information technology, 2003. 22(1): p. 1-16.
25. Wu, J.-T.B. and G.M. Marakas, The impact of operational user participation on perceived system implementation success: An empirical investigation. Journal of Computer Information Systems, 2006. 46(5): p. 127-140.
26. Myers, M.D., Qualitative research in information systems. Management Information Systems Quarterly, 1997. 21(2): p. 241-242.
27. King, N., C. Cassell, and G. Symon, Using templates in the thematic analysis of text. Essential guide to qualitative methods in organizational research, 2004. 2: p. 256-70.
28. Tan, F.B. and M.G. Hunter, The repertory grid technique: A method for the study of cognition in information systems. Mis Quarterly, 2002: p. 39-57.
29. Curtis, A.M., et al., An overview and tutorial of the repertory grid technique in information systems research. 2008.
30. Runeson, P. and M. Höst, Guidelines for conducting and reporting case study research in software engineering. Empirical software engineering, 2009. 14(2): p. 131.

## APPENDIX A: Questionnaire for Interviews

DEVELOPMENT TEAM MEMBERS

**Background Information**
- Please briefly describe the aim of the software development project you are currently involved in.
- Please describe the software development process your team is using to manage this project? Would you describe it as agile?
- How would you describe the current stage of the project?
- Please describe your role in this project.
- What do you consider your main responsibilities in this role for this project?
- How many years of experience do you have in this role?
- How many years have you been with this organisation?

**Characteristics of users to be involved**
- Who do you consider are the main client stakeholders that need to be involved in this project and what are your expectations about their knowledge and skills?
- How will you go about selecting users for participation in project related activities?
- What characteristics, according to you, make a good candidate for participation in project related activities?

**Expectations of User Involvement**
- How do you think the interaction with the users will contribute to this project? / What are you hoping to get out of the interaction with users?
- In what ways do you think users of the software should be involved in the project and how will you go about getting them to contribute to the project? (stages of development, activities, expectations of time commitments)
- In your experience, how aligned were the users' expectations of their role in the project with yours? In what way does this alignment or misalignment affect the success of the involvement and by extension, the project?

**Defining the Quality of User Involvement**
- How would you judge the effectiveness of users' participation in/contribution to the project?
- How do you distinguish between useful user contribution and poor user contribution? (Probe)
- What are the barriers expected when involving users in project activities and how would you go about alleviating them?
- If the involvement does yield the desired results, what will you do to get optimal benefit from the participation of users?

STAKEHOLDER TEAM MEMBERS

**Background Information**
- Can you please describe your role in in the organization? • How many years' experience do you have in this role?
- How many years have you been with this organization?
- Have you been involved in a development project before?

**Characteristics of user involvement**
- What knowledge or skills do you think you will be expected to have during your involvement in this project?
- User Involvement
- Please briefly describe the aim of the software development
- project you are currently involved in.
- How did you come about to participate in the project?
- What do you know about what you will be doing in the project and how do you think you will be contributing to this project?
- About how much time do you think you will have to invest through the duration of this project and what are your thoughts on the expectations/demands of your time?
- What do you think can be done to improve your understanding of your role in this project and help you contribute effectively?
- Who do you think you will be interacting with from the development team (the roles)?
- What are your thoughts on being part of this project?

**Defining the Quality of User Involvement**
- How do you think the effectiveness of your contribution will be judged?
- What according to you differentiates effective participation from ineffective participation in project related activities (assigned to you)